\documentclass[conference]{IEEEtran}
\usepackage{amsmath,amsopn,amssymb,bm}
\usepackage{amsfonts}
\usepackage{gensymb}
\usepackage{amsthm} 
\usepackage{graphicx,xspace,color}
\usepackage{epsfig}
\usepackage{longtable,multirow}
\usepackage{array}
\usepackage{stfloats}
\usepackage{cuted}
\usepackage{cite}
\usepackage[nolist]{acronym}
\usepackage{soul}
\usepackage{algorithm,algpseudocode}
\usepackage{algcompatible}
\usepackage{enumerate}
\usepackage{lettrine}
\usepackage{mathtools}
\usepackage{comment}
\usepackage{subcaption}

\usepackage[font=footnotesize]{caption}

\usepackage{booktabs}
\graphicspath{ {./figures/}}

\newtheorem{remark}{Remark}
\newtheorem{proposition}{Proposition}

\begin{acronym}
	\acro{THz}{Terahertz}
	\acro{IRS}{intelligent reflecting surface}
	\acro{RF}{radio-frequency}
	\acro{SNR}{receive signal-to-noise ratio}
	\acro{SW}{spherical wave}
	\acro{MIMO}{multiple-input multiple-output}
	\acro{EE}{energy efficiency}
	\acro{UPA}{uniform planar array}
	\acro{LoS}{line-of-sight}
\end{acronym}	

\begin{document}
	
\title{Intelligent Reflecting Surfaces at Terahertz Bands: Channel Modeling and Analysis}
	\author{\IEEEauthorblockN{Konstantinos Dovelos\IEEEauthorrefmark{1}, Stylianos D. Assimonis\IEEEauthorrefmark{2}, Hien Quoc Ngo\IEEEauthorrefmark{2}, Boris Bellalta\IEEEauthorrefmark{1}, and Michail Matthaiou\IEEEauthorrefmark{2}}
	\IEEEauthorblockA{$^*$Department of Information and Communication Technologies, Universitat Pompeu Fabra (UPF), Barcelona, Spain}
	\IEEEauthorblockA{$^\dagger$Institute of Electronics, Communications and Information Technology (ECIT), Queen's University Belfast, Belfast, U.K.}
	Email: \{konstantinos.dovelos, boris.bellalta\}@upf.edu, \{s.assimonis, hien.ngo, m.matthaiou\}@qub.ac.uk}
\maketitle

\begin{abstract}
An intelligent reflecting surface (IRS) at terahertz (THz) bands is expected to have a massive number of reflecting elements to compensate for the severe propagation losses. However, as the IRS size grows, the conventional far-field assumption starts becoming invalid and the spherical wavefront of the radiated waves should be taken into account. In this work, we consider a spherical wave channel model and pursue a comprehensive study of IRS-aided multiple-input multiple-output (MIMO) in terms of power gain and energy efficiency (EE). Specifically, we first analyze the power gain under beamfocusing and beamforming, and show that the latter is suboptimal even for multiple meters away from the IRS. To this end, we derive an approximate, yet accurate, closed-form expression for the loss in the power gain under beamforming. Building on the derived model, we next show that an IRS can significantly improve the EE of  MIMO when it operates in the radiating near-field and performs beamfocusing. Numerical results corroborate our analysis and provide novel insights into the design and performance of IRS-assisted THz~communication. 

\end{abstract}

\section{Introduction}
\ac{THz} communication is widely deemed a key enabler for future 6G wireless networks due to the abundance of available spectrum at \ac{THz} bands~\cite{6G_survey}. However, \ac{THz} wireless links are subject to severe propagation losses, which require transceivers with a massive number of antennas to compensate for these losses and extend the communication range~\cite{5G_prospective}. On the other hand, unlike sub-6 GHz systems, the power consumption of \ac{THz} \ac{RF} circuitry is considerably high, which might undermine the deployment of large antenna arrays in an energy efficient manner~\cite{aosa_thz}. To overcome this problem, the novel concept of \ac{IRS} can be exploited to build transceivers with a relatively small number of antennas, which work along with an \ac{IRS} to achieve high spectral efficiency with reduced power consumption~\cite{IRS_commun}. Thus, the performance analysis of \ac{IRS}-aided \ac{THz} communication is of great research importance. 

There is a large body of literature investigating the modeling and performance of \ac{IRS}s at sub-6 GHz and millimeter wave bands. Most of them, though, treat the \ac{IRS} element as a typical antenna that re-radiates the impinging wave, and leverage antenna theory to characterize the path loss of the \ac{IRS}-aided link. Furthermore, they assume far-field, where the spherical wavefront of the emitted waves degenerates to a plane wavefront. Although these approaches are popular due to their simplicity, they might not capture the unique features of \ac{IRS}s, and especially at \ac{THz} bands. To this direction,~\cite{pathloss_model1} introduced a path loss model for the sub-6 GHz band by invoking \textit{plate scattering theory}, but assuming a specific scattering plane; hence, it is applicable only to special cases. The authors in~\cite{pathloss_model2} extended the said path loss model to arbitrary incident angles and polarizations, but considered the far-field zone of the \ac{IRS}. Recently, a stream of papers (e.g., \cite{nf_pathloss_model1}, \cite{nf_pathloss_model2}, and references therein) proposed a path loss model that is applicable to near-field, using the ``cos$^q$'' radiation pattern~\cite{reflectarray_book} for each \ac{IRS} element, which differs from the plate scattering-based model.

Although there are still many critical questions about the operation of \ac{THz} IRSs, there is a dearth of related literature. From related work, we distinguish~\cite{thz_holographic_irs}, where the authors showed that the far-field beampattern of a holographic \ac{IRS} can be well approximated by that of an ultra-dense \ac{IRS}, and then proposed a channel estimation scheme for \ac{THz} massive \ac{MIMO} aided by a holographic \ac{IRS}. However, due to the high propagation losses and the short wavelength, a \ac{THz}  \ac{IRS} is expected to consist of a massive number of passive reflecting elements, resulting in a radiating near-field, i.e., Fresnel zone, of several meters. Additionally, to effectively overcome the path loss of the transmitter-\ac{IRS} link, the transmitter will need to operate near the \ac{IRS}, which is in sharp contrast to sub-6 GHz massive \ac{MIMO} of macrocell deployments. In conclusion, a \ac{THz} \ac{IRS} calls for a carefully tailored design that takes into account the aforementioned particularities. 

This paper aims to shed light on these aspects, and study the channel modeling and performance of \ac{THz} \ac{IRS}. In particular:
\begin{itemize}
\item We provide a near-field channel model for \ac{THz} frequencies. Our model is physically consistent, and takes into account the size of the \ac{IRS} elements in the path loss calculation, as well as in the spherical wavefront of the radiated waves.  
\item We show that a typical \ac{THz} \ac{IRS} is likely to operate in the Fresnel zone, where conventional beamforming is suboptimal and hence can reduce the power gain. More importantly, we analytically evaluate that loss by providing an approximate closed-form expression. 
\item We compare \ac{IRS}-aided \ac{MIMO} with \ac{MIMO}, and demonstrate the \ac{EE} gains of the former architecture. More specifically, we determine the optimal number of \ac{IRS} elements required to attain the same rate as \ac{MIMO} with reduced power consumption, and reveal the \ac{EE} scaling laws. 
\end{itemize}

\textit{Notation}: $D_{N}(x) = \frac{\sin(Nx/2)}{N\sin(x/2)}$ is the Dirichlet sinc function; $\mathbf{A}$ is a matrix; $\mathbf{a}$ is a vector; $[\mathbf{A}]_{i,j}$ is the $(i,j)$th entry of $\mathbf{A}$; $\mathbf{A}^T$ and $\mathbf{A}^H$ are the transpose and conjugate transpose of $\mathbf{A}$, respectively, $\text{vec}(\mathbf{A})$ is the column vector formed by stacking the columns of $\mathbf{A}$; and $\mathcal{CN}(\bm{\mu},\bm{\Sigma})$ is a complex Gaussian vector with mean $\bm{\mu}$ and covariance matrix $\bm{\Sigma}$.
\begin{figure}[t]
	\centering
	\includegraphics[width=1\linewidth]{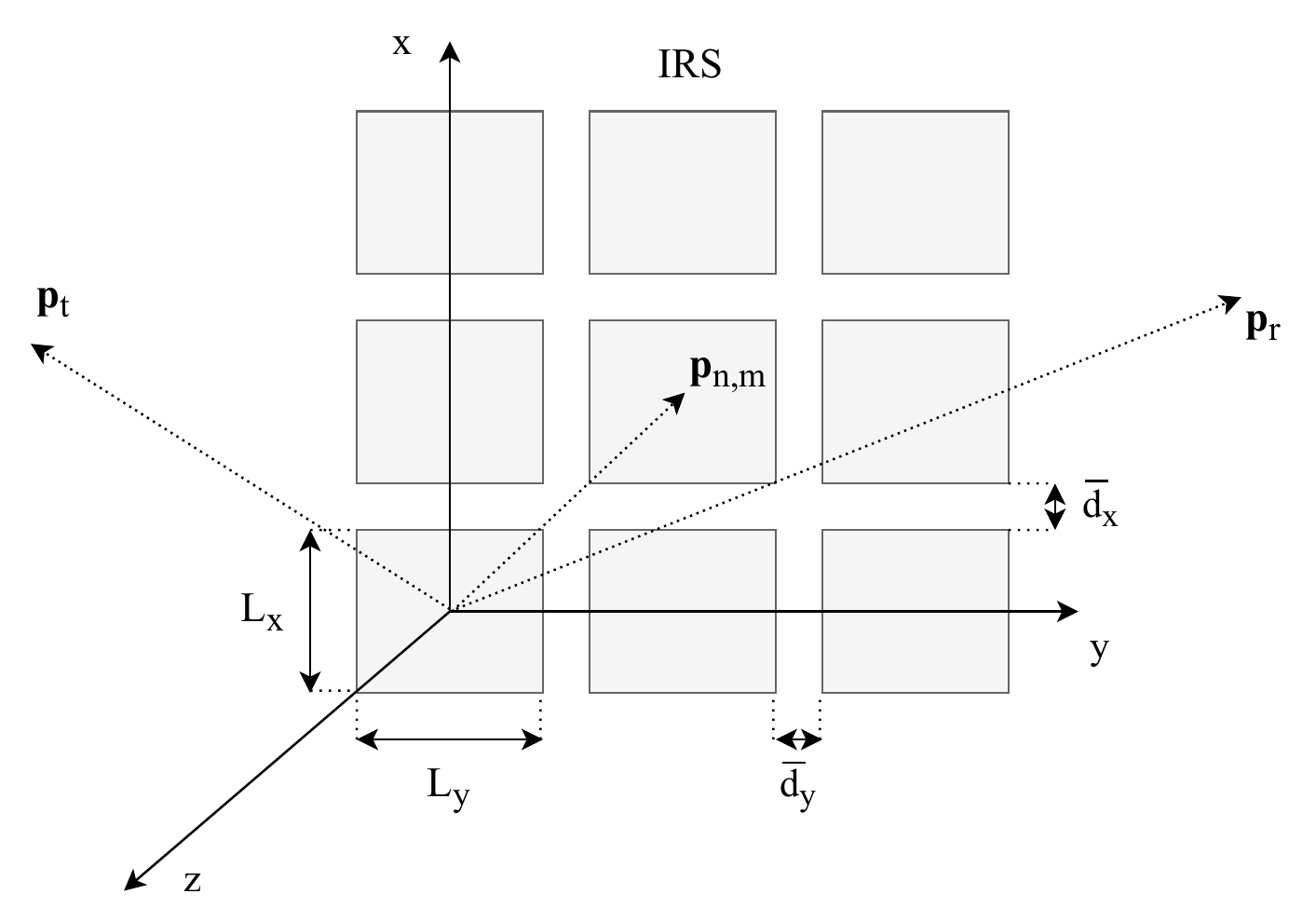}
	\caption{Illustration of the \ac{IRS} geometry considered in the channel model.}
	\label{fig:Fig_irs}
\end{figure}
\section{System Model}
\subsection{Signal Model} Consider a \ac{THz} IRS system where the transmitter (Tx) and receiver (Rx) have a single antenna each. The IRS is placed in the $xy$-plane, and it consists of $N = N_x\times N_y$ passive reflecting elements. Each reflecting element is of size $L_x\times L_y$, and the spacings between adjacent elements are $\bar{d}_x$ and $\bar{d}_y$ along the $x$ and $y$ directions, respectively. The reflection coefficient of the $(n,m)$th \ac{IRS} element is $e^{j\varphi_{n,m}}$, where $\varphi_{n,m}\in[-\pi, \pi]$. We next focus on the Tx-IRS-Rx link. The baseband signal at the receiver is then written as
\begin{equation}\label{eq:rx_signal}
y = \mathbf{h}^T_r\mathbf{\Phi}\mathbf{h}_t s+ \tilde{n},
\end{equation}
where $\mathbf{\Phi} = \text{diag}(\text{vec}(\tilde{\mathbf{\Phi}}))\in\mathbb{C}^{N\times N}$, with $\tilde{\mathbf{\Phi}}\in\mathbb{C}^{N_x\times N_y}$ and $[\tilde{\mathbf{\Phi}}]_{n,m} = e^{j\varphi_{n,m}}$, is the \ac{IRS}'s reflection matrix, $\mathbf{h}_r\in\mathbb{C}^{N\times 1}$ is the channel from the Rx to the IRS, $\mathbf{h}_t\in\mathbb{C}^{N\times 1}$ is the channel from the Tx to the IRS, $s\sim\mathcal{CN}(0,P_t)$ is the transmitted data symbol, $P_t$ is the average power per data symbol, and $\tilde{n}\sim\mathcal{CN}(0,\sigma^2)$ is the additive noise. 

\subsection{Channel Model}
\subsubsection{Spherical Wavefront} Unlike antenna arrays that are typically modeled as a collection of point radiators, an IRS comprises rectangular reflecting elements whose size cannot be neglected. Assume that the center of the $(0,0)$th IRS element is placed at the origin of the coordinate system, as shown in Fig.~\ref{fig:Fig_irs}. Across the $(n,m)$th \ac{IRS} element, the reflection coefficient $e^{j\varphi_{n,m}}$ remains constant, and the phase difference between adjacent elements is measured from their centers. Thus, the position vector of the $(n,m)$th IRS element is $\mathbf{p}_{n,m} = \left(nd_x, md_y, 0\right)$, where $d_x \triangleq \bar{d}_x + L_x$ and $d_y \triangleq \bar{d}_y + L_y$. Let $\lambda$ denote the carrier wavelength. Henceforth, we consider $L_x = L_y  = \lambda/2$ and $\bar{d}_x=\bar{d}_y=0$~\cite{nf_pathloss_model2,em_perspective}. 

The Tx and Rx are located in $(D_t,\theta_t,\phi_t)$ and $(D_r,\theta_r,\phi_r)$, respectively, and hence their position vectors in Cartesian coordinates are
\begin{align}
\mathbf{p}_t & = (D_t\cos\phi_t\sin\theta_t, D_t\sin\phi_t\sin\theta_t, D_t\cos\theta_t), \\
\mathbf{p}_r & = (D_r\cos\phi_r\sin\theta_r, D_r\sin\phi_r\sin\theta_r, D_r\cos\theta_r),
\end{align}
where $D_t$ and $D_r$ are the distances measured from the (0,0)th IRS element, while $\phi$ and $\theta$ denote the azimuth and polar angles, respectively. The baseband channel from the Tx to the IRS is specified as $\mathbf{h}_t = \text{vec}(\mathbf{M}_t)$, where $\mathbf{M}_t\in\mathbb{C}^{N_x\times N_y}$ is the auxiliary matrix with entries~\cite{position_sw}
\begin{equation}\label{eq:ch_matrix_tx}
[\mathbf{M}_t]_{n,m} = \sqrt{\text{PL}^{t}_{n,m}}e^{-j k D^{t}_{n,m}}.
\end{equation}
In~\eqref{eq:ch_matrix_tx}, $k = \frac{2\pi}{\lambda}$ is the wavenumber, $\text{PL}^{t}_{n,m}$ is the path loss between the Tx and the $(n,m)$th IRS element, and $D^t_{n,m}\triangleq \| \mathbf{p}_t - \mathbf{p}_{n,m}\|$ is the respective distance, with $D^t_{0,0} = D_t$. Similarly, we have $\mathbf{h}_r = \text{vec}(\mathbf{M}_r)$, where $\mathbf{M}_r\in\mathbb{C}^{N_x\times N_y}$ is the auxiliary matrix with entries
\begin{equation}\label{eq:ch_matrix_rx}
[\mathbf{M}_r]_{n,m} = \sqrt{\text{PL}^{r}_{n,m}}e^{-jk D^{r}_{n,m}},
\end{equation}
where $\text{PL}^{r}_{n,m}$ is the path loss between the Rx and the $(n,m)$th IRS element, and $D^r_{n,m}\triangleq \| \mathbf{p}_r - \mathbf{p}_{n,m}\|$, with $D^r_{0,0} = D_r$. Using~\eqref{eq:ch_matrix_tx} and~\eqref{eq:ch_matrix_rx}, the received signal in \eqref{eq:rx_signal} is recast as 
\begin{align}
y = \sum_{n=0}^{N_x-1} \sum_{m=0}^{N_y-1}\sqrt{\text{PL}_{n,m}}e^{-j k(D^{t}_{n,m} + D^{r}_{n,m})}e^{j\varphi_{n,m}} s+ \tilde{n},
\end{align}
where $\text{PL}_{n,m} = \text{PL}^{t}_{n,m}\text{PL}^{r}_{n,m}$ denotes the overall path loss of the Tx-IRS-Rx link through the $(n,m)$th reflecting element. Hence, the \ac{SNR} is 
\begin{equation}\label{eq:receive_snr}
\text{SNR} = \frac{P_t}{\sigma^2} \left|\sum_{n=0}^{N_x-1} \sum_{m=0}^{N_y-1}\sqrt{\text{PL}_{n,m}}e^{-j k(D^{t}_{n,m} + D^{r}_{n,m})}e^{j\varphi_{n,m}}\right|^2.
\end{equation}
In the sequel, we detail the path loss model for \ac{THz} bands, which relies on the plate scattering paradigm~\cite{balanis_book}.
\subsubsection{Scattered Field by an IRS Element} We focus on an arbitrary \ac{IRS} element and omit the subscript ``$n,m$'' hereafter. The Tx and Rx are in the far-field of the \textit{individual element}, which implies that $D_t, D_r >2L^2_{\max}/\lambda$, where $2L^2_{\max}/\lambda$ is the Fraunhofer distance and $L_{\max} = \max(L_x,L_y)$ is the maximum dimension of the element. Consequently, a plane wavefront is assumed across the \ac{IRS} element. For simplicity, we consider a transverse electric incident wave which is linearly polarized along the $x$-axis. The electric field (E-field) of the incident plane wave is hence given by 
\begin{align}
\mathbf{E}_{i} &= E_i e^{-j k (y\sin\theta_t - z\cos\theta_t)}\mathbf{e}_x,
\end{align}
where $\mathbf{e}_x$ denotes the unit vector along the $x$-axis. Next, the scattered field $\mathbf{E}_s$ at the receiver location $(D_r,\theta_r,\phi_r)$ is determined using physical optics techniques, whereby the IRS element is modeled as a perfectly conducting plate. Specifically, the squared magnitude of the scattered E-field\footnote{The \ac{IRS} elements can alter the phase of the scattered wave. The reflection coefficient does not appear in the formula of $\|\mathbf{E}_s\|^2$ since $|e^{j\varphi_{n,m}}|^2 = 1$.} is given by~\cite[Ch.~11]{balanis_book}
\begin{align}
\|\mathbf{E}_s\|^2 &=  \left(\frac{L_xL_y}{\lambda}\right)^2\frac{|E_i|^2}{D^2_r}F(\theta_t,\phi_r,\theta_r)\text{sinc}^2(X)\text{sinc}^2(Y) \\
&\approx \left(\frac{L_xL_y}{\lambda}\right)^2\frac{|E_i|^2}{D^2_r}F(\theta_t,\phi_r,\theta_r),\label{eq:Es_approx}
\end{align}
where $F(\theta_t,\phi_r,\theta_r) \triangleq \cos^2\theta_t(\cos^2\theta_r\cos^2\phi_r + \sin^2\phi_r)$, while $X\triangleq \frac{\pi L_x}{\lambda}\sin\theta_r\cos\phi_r$ and $Y\triangleq \frac{\pi L_y}{\lambda}(\sin\theta_r\sin\phi_r - \sin\theta_t)$. 
\begin{figure}[t]
	\centering
	\includegraphics[width=0.79\linewidth]{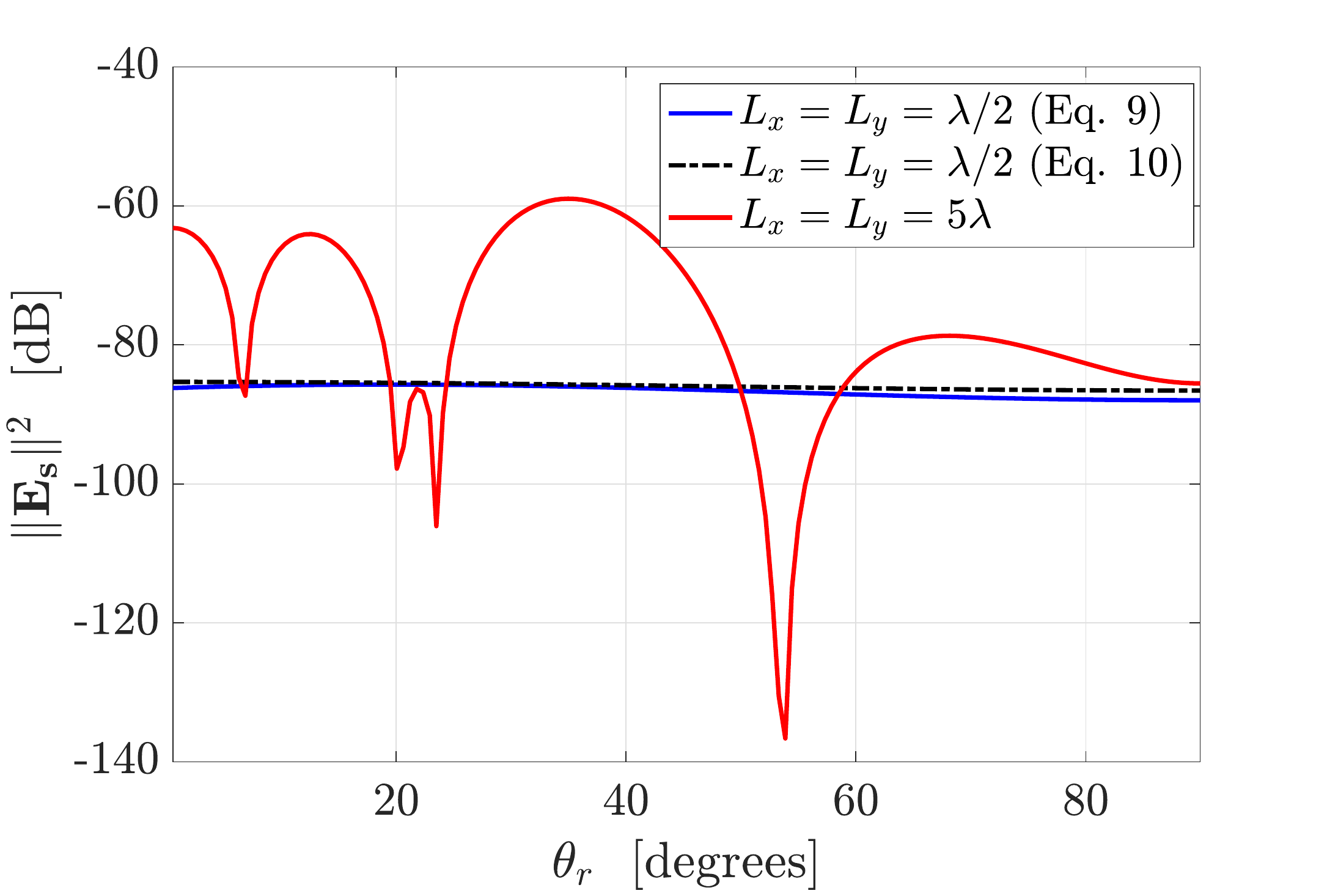}
	\caption{Squared magnitude of the scattered field versus observation angle $\theta_r$ for incident angle $\theta_t = 30\degree$ and scattering plane $\phi_r = 60\degree$; $|E_i|^2=1$, carrier frequency $ f = 300$ GHz, and $D_r = 4$ meters.}
	\label{fig:Fig_scatteredField}
\end{figure}
The approximation in~\eqref{eq:Es_approx} follows from $\text{sinc}(X)\approx 1$ and $\text{sinc}(Y)\approx 1$ for $X\approx 0$ and $Y\approx 0$, which holds for $L_x \leq \lambda$ and $L_y \leq \lambda$. This is also verified in Fig.~\ref{fig:Fig_scatteredField}. It is worth stressing that each IRS element is expected to be of sub-wavelength size in order to act as an \text{isotropic scatterer}~\cite{smart_radio}.
\subsubsection{Path Loss} 
\begin{figure}[t]
	\centering
	\includegraphics[width=0.77\linewidth]{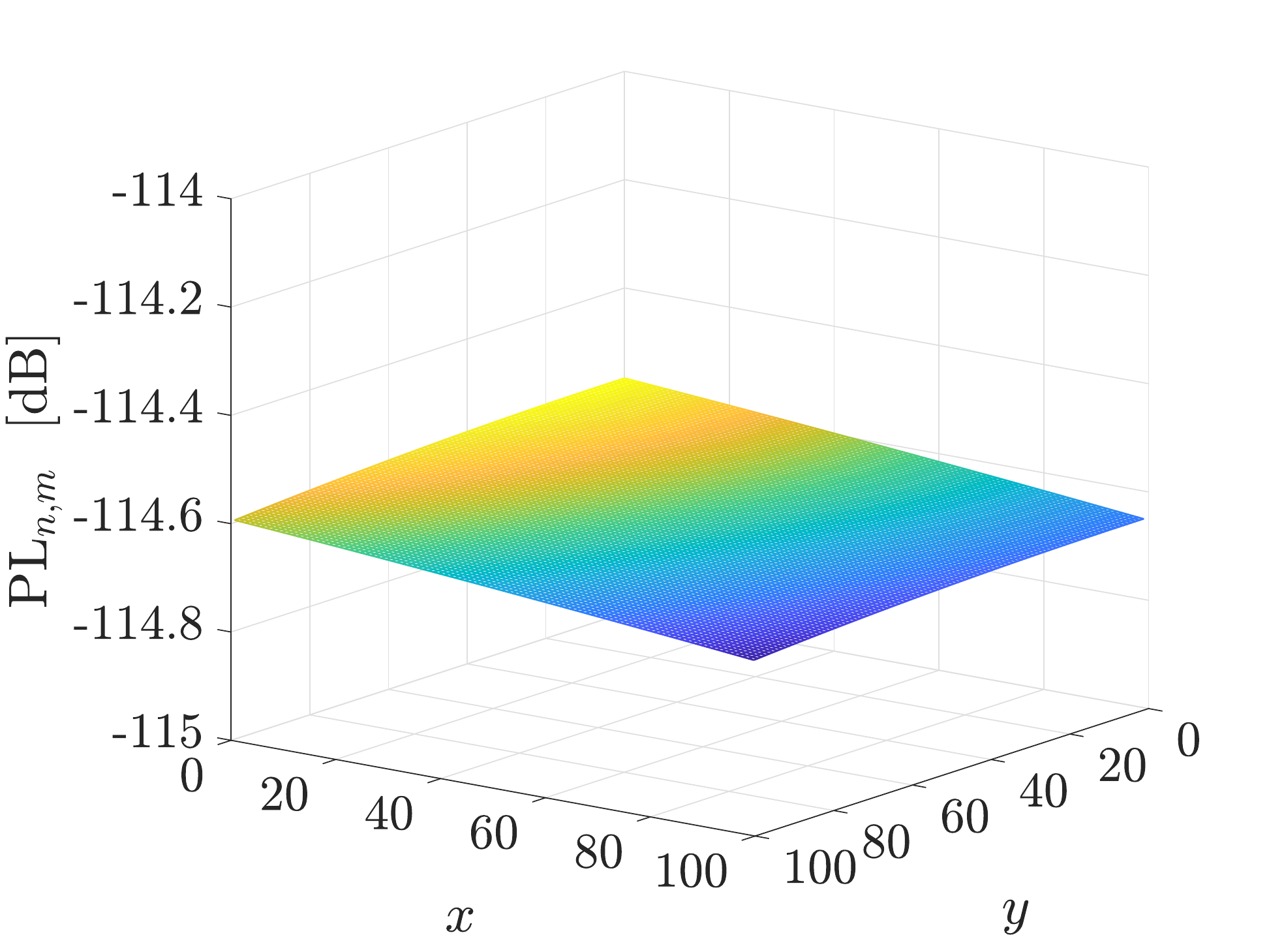}
	\caption{Path loss across an $100\times 100$-element IRS; $f = 300$~GHz, $\kappa_{\text{abs}}(f) = 0.0033$ m$^{-1}$, $L_x = L_y = \lambda/2$, $\bar{d}_x=\bar{d}_y=0$, $G_t = G_r = 20$ dBi, $\mathbf{p}_t = (0, -0.3, 0.6)$, $D_t = 0.67$ m, $\mathbf{p}_r = (0, 1, 1)$, and $D_r = 1.41$ m.}
	\label{fig:Fig_PL}
\end{figure}

Recall that the relation between $P_t$ and $E_i$ is $|E_i|^2/(2\eta) = P_tG_t/(4\pi D^2_t)$, where $\eta$ is the free-space impedance, and $G_t$ is the transmit antenna gain~\cite{balanis_book2}. Hence, the power density of the scattered field is 
\begin{equation}
S_s = \frac{\|\mathbf{E}_s\|^2}{2\eta} = \left(\frac{L_xL_y}{\lambda}\right)^2\frac{P_tG_t}{4\pi D_t^2D^2_r}F(\theta_t,\phi_r,\theta_r).
\end{equation}
Considering the receive aperture $A_r  = G_r \lambda^2/(4\pi)$ yields the receive power 
\begin{equation}\label{eq:rx_power }
P_r = S_s A_r =  P_t \frac{G_tG_r}{(4\pi D_t D_r)^2}(L_xL_y)^2F(\theta_t,\phi_r,\theta_r). 
\end{equation}
Finally, taking into account the molecular absorption losses at \ac{THz} bands gives the path loss of the Tx-IRS-Rx link through the $(n,m)$th element
\begin{align}\label{eq:PL_nm}
\text{PL}_{n,m} =  \frac{G_tG_r(L_xL_y)^2}{(4\pi D^t_{n,m} D^r_{n,m})^2}F(\theta_t,\phi_r,\theta_r)e^{-\kappa_{\text{abs}}(f)(D^t_{n,m} + D^r_{n,m})},
\end{align}
where $\kappa_{\text{abs}}(f)$ is the molecular absorption coefficient at the carrier frequency $f$. From Fig.~\ref{fig:Fig_PL}, we see that $\text{PL}_{n,m}$ marginally changes across the \ac{IRS}, even for $100\times 100$ elements and a Tx distance $D_t = 0.67$ m. This is because of the small physical size of the \ac{IRS} at \ac{THz} bands. Hereafter, we will assume that $\text{PL}_{n,m}\approx \text{PL}$, where $\text{PL}$ is calculated using $D_t$ and $D_r$ measured from the $(0,0)$th IRS element.  
\begin{table}[H]
	\centering
	\caption{IRS with $L_x=L_y = \lambda/2$ and $\bar{d}_x=\bar{d}_y=0$ at $f=300$ GHz.}
	\label{table:key_params}
	\begin{tabular}{|c  c  c|}
		\hline
		$N_x\times N_y$-elements  & Physical Size [$\text{m}^2$] &  Fresnel Region [m]\\
		\hline
		$80\times 80$ & $0.039 \times 0.039$ & $[0.15, 3.2]$ \\
		\hline
		$100\times 100$ & $0.05 \times 0.05$ & $[0.22, 5]$ \\
		\hline
	\end{tabular}
\end{table}
\section{Power Gain of IRS-Aided THz System}
\subsection{Fresnel Region}
The near-field of an IRS refers to distances that are smaller than the \text{Fraunhofer distance} $D_F \triangleq 2 L^2_{\text{IRS}}/\lambda$, where $L_{\text{IRS}}\triangleq\max\left(N_xL_x + (N_x-1)\bar{d}_x, N_yL_y + (N_y-1)\bar{d}_y\right)$ is the maximum physical dimension of the IRS. In our work, we focus on the radiating near-field, i.e., Fresnel region, which corresponds to distances $D \gg \lambda$ satisfying~\cite{balanis_book2}
\begin{equation}
0.62 \sqrt{L^3_{\text{IRS}}/\lambda} < D \leq 2 L^2_{\text{IRS}}/\lambda.
\end{equation}
 From Table~\ref{table:key_params}, we verify the small physical size of THz IRS, as well as its large Fresnel region. Consequently, it is very likely that the Tx and Rx are in the near-field of the IRS, where the spherical wavefront of the impinging waves \textit{across the IRS} cannot be neglected.
\subsection{Near-Field Beamfocusing}
Let us define the normalized power gain as
\begin{equation}\label{eq:normalized_power_gain}
G \triangleq \frac{\left|\sum_{n=0}^{N_x-1}\sum_{m=0}^{N_y-1}e^{-jk\left(D^t_{n,m}  + D^r_{n,m}\right)}e^{j\varphi_{n,m}}\right|^2}{N_x^2N_y^2},
\end{equation}
with $G\in[0,1]$. The receive \ac{SNR} in~\eqref{eq:receive_snr} is now written as 
\begin{equation}
\text{SNR} \approx \frac{N^2GP_t \text{PL}}{\sigma^2}.
\end{equation}
The power gain is maximized by near-field beamfocusing. Hence, the phase induced by the $(n,m)$th IRS element is 
\begin{equation}\label{eq:beamfocusing}
\varphi_{n,m} = k\left(D^t_{n,m}  + D^r_{n,m}\right),
\end{equation}
which yields $G=1$ and $\text{SNR}=N^2P_t \text{PL}/\sigma^2$. As expected, the \ac{SNR} of an IRS-aided system grows quadratically with the number $N$ of IRS elements~\cite{passive_and_active_bf}. Note, though, that the \ac{IRS} needs to know the exact locations of the Tx and Rx in order to perform beamfocusing. 
 \begin{figure*}[t]
	\centering
	\begin{subfigure}{.5\textwidth}
		\centering
		\includegraphics[width=.72\linewidth]{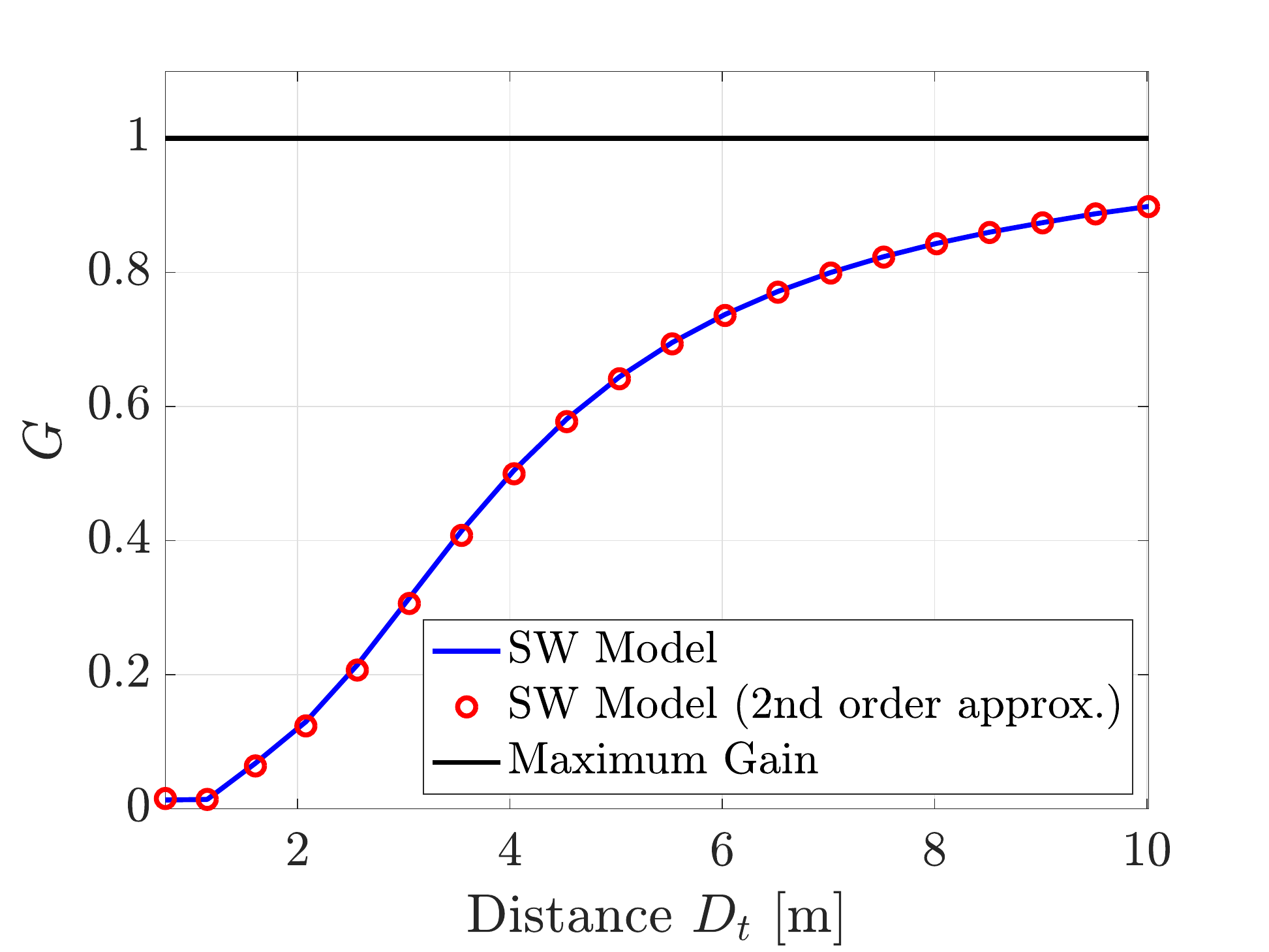}
		\caption{}
		\label{fig:Fig0_a}
	\end{subfigure}%
	\begin{subfigure}{.5\textwidth}
		\centering
		\includegraphics[width=.72\linewidth]{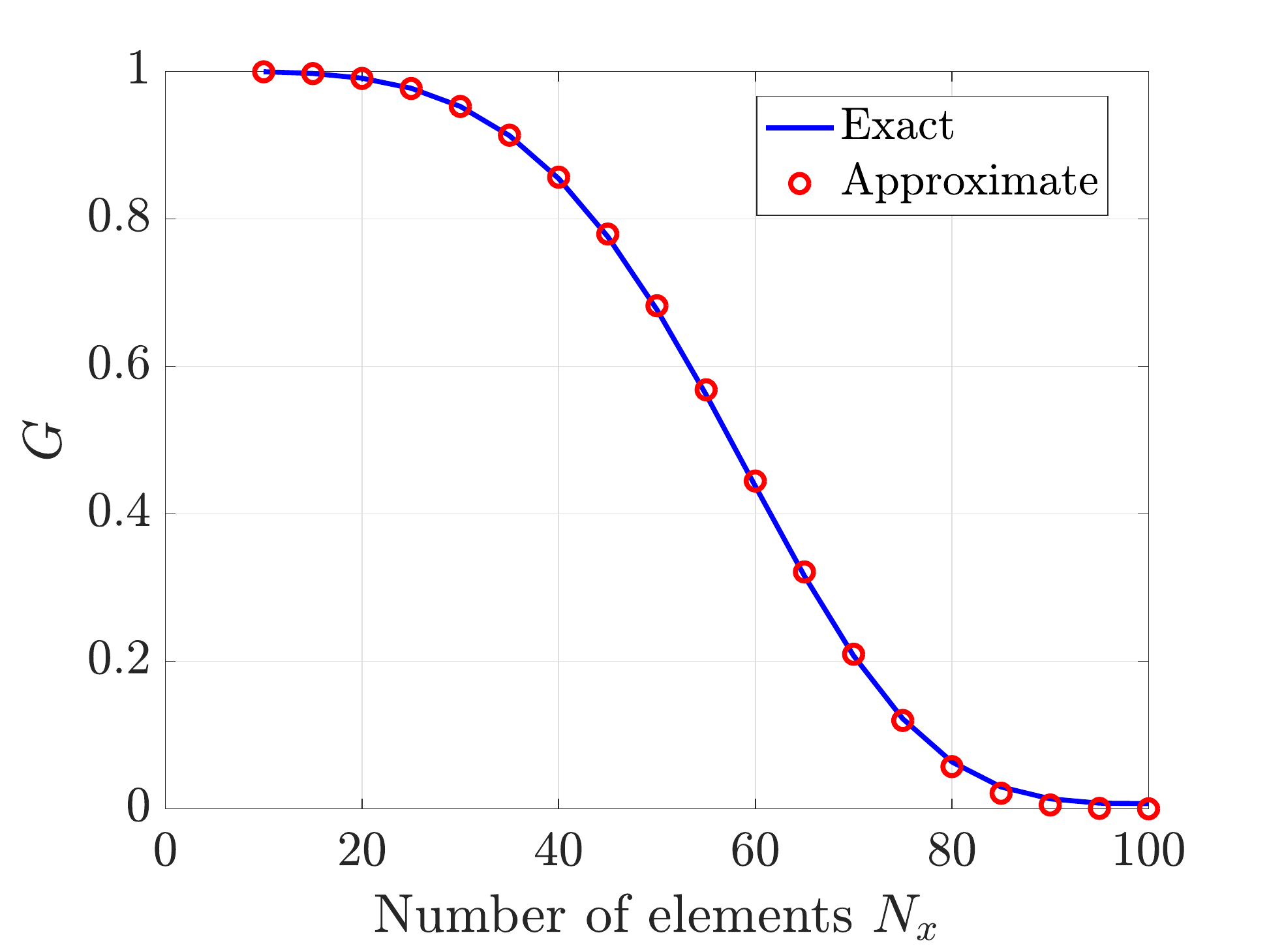}
		\caption{}
		\label{fig:Fig0_b}
	\end{subfigure}
	\caption{(a) Normalized power gain vs. distance $D_t$ for an $100\times 100$-element IRS, where $\mathbf{p}_t = (0.4,0.4, z)$, $ 0.5 \leq z \leq 10$, and $0.755  \leq D_t \leq 10.016$~m. (b) Normalized power gain vs. number of elements for an $N_x\times N_y$-element IRS, where $N_x=N_y$, $\mathbf{p}_t = (0.4, 0.4, 1)$, and $ D_t =1.15$ m. The other parameters are $f = 300$ GHz, $L_x=L_y=\lambda/2$, and  $\bar{d}_x=\bar{d}_y=0$.}
	\label{fig:Fig0}
\end{figure*}

\subsection{Far-Field Beamforming}
In this section, we analyze the power gain under conventional far-field beamforming, which relies on the parallel ray approximation. First, using basic  algebra, we have that 
\begin{align}\label{eq:sw_model}
D^{r}_{n,m}  = &D_r\left(1 + \frac{(nd_x)^2}{D_r^2} - \frac{2\cos\phi_r\sin\theta_r nd_x}{D_r} \right . \nonumber\\
 & \quad  \quad \quad \left . + \frac{(md_y)^2}{D_r^2} - \frac{2\sin\phi_r\sin\theta_r md_y}{D_r}\right)^{1/2}.
\end{align}
In the far-field $D_r\gg D_F$, the first-order Taylor expansion $(1 + x)^a \approx 1 + a x$ can be applied to~\eqref{eq:sw_model}, while ignoring the quadratic terms $(nd_x)^2/D_r^2$ and $(md_y)^2/D_r^2$. This yields
\begin{equation}\label{eq:pw_model}
D^r_{n,m} \approx D_r- nd_x\cos\phi_r\sin\theta_r -md_y\sin\phi_r\sin\theta_r,
\end{equation}
which corresponds to the plane wavefront model. 
\begin{remark}\label{remark1}
The far-field steering vector is defined as $\mathbf{a}(\phi,\theta) \triangleq \emph{vec}(\mathbf{M})$, where $\mathbf{M}\in\mathbb{C}^{N_x\times N_y}$ is the matrix with elements $[\mathbf{M}]_{n,m} = e^{jk (nd_x\cos\phi \sin\theta  + md_y\sin\phi \sin\theta)}$. Thus, the channel vector is $\mathbf{h} =  \sqrt{\emph{PL}}e^{-jk D} \mathbf{a}(\phi,\theta)$. 
\end{remark}
Let us now consider that the Rx is in the far-field of the IRS whilst the Tx is close to the IRS; in fact, this deployment yields the maximum \ac{SNR}, compared to placing the \ac{IRS} somewhere in between~\cite{tutorial}. If the IRS employs beamforming based on the angular information $(\phi_t,\theta_t)$ and $(\phi_r,\theta_r)$, i.e., 
\begin{align}\label{eq:bf_strategy}
\varphi_{n,m} = - k(&nd_x\cos\phi_t\sin\theta_t + md_y\sin\phi_t\sin\theta_t\nonumber \\
& + nd_x\cos\phi_r\sin\theta_r +md_y\sin\phi_r\sin\theta_r),
\end{align}
the power gain will  decrease. To analytically characterize this reduction, we use the second-order Taylor expansion $(1 + x)^a \approx 1 + ax + \frac{1}{2}a(a-1) x^2$ and neglect the terms $O(d^q/D^q), q\geq~3$, which yields the (Fresnel) approximation of the Tx distance
\begin{align}\label{eq:fresnel_approximation}
D^{t}_{n,m} \approx D_t  &+ \frac{(nd_x)^2(1-\cos^2\phi_t\sin^2\theta_t)}{2D_t} - nd_x\cos\phi_t \sin\theta_t \nonumber\\
&+ \frac{(md_y)^2(1-\sin^2\phi_t\sin^2\theta_t)}{2D_t} - md_y\sin\phi_t\sin\theta_t.
\end{align}  
Using~\eqref{eq:pw_model},~\eqref{eq:bf_strategy} and~\eqref{eq:fresnel_approximation}, the normalized power gain in~\eqref{eq:normalized_power_gain} reduces to the expession~\eqref{eq:losses} at the top of the last page. The accuracy of the approximation of the Tx distance is depicted in~Fig.~\ref{fig:Fig0}(\subref{fig:Fig0_a}), and the validity of~\eqref{eq:losses} is evaluated in~Fig.~\ref{fig:Fig0}(\subref{fig:Fig0_b}). Note that the lower limit of the Fresnel zone of an $100\times 100$-element IRS, with $L_x=L_y=\lambda/2$ and $\bar{d}_x=\bar{d}_y=0$, is $0.22$~meters according to Table~I. Thus, the distances in the numerical experiments were chosen so that the Tx does not operate in the reactive near-field. As observed, beamforming can substantially decrease the power gain even for distances of several meters away from the \ac{IRS}. This is because of the mismatch between~\eqref{eq:beamfocusing} and~\eqref{eq:bf_strategy}. Moreover, from~\eqref{eq:losses}, we have the asymptotic result $G\to 0$ as $N\to\infty$. In conclusion, near-field beamfocusing should be used in most cases of interest.

\section{Performance of IRS-Aided THz System}\label{sec:perf_analysis}

\subsection{Benchmark: MIMO System}
Consider a \ac{MIMO} system, where the Tx and Rx are equipped with $N_t$ and $N_r$ antennas, respectively. For efficient hardware implementation, hybrid array architectures are assumed at both ends. The path loss of the direct channel, i.e., \ac{LoS}, is given by
\begin{equation}\tag{23}
\text{PL}_{\text{MIMO}} = \frac{G_tG_r\lambda^2}{(4\pi D_d)^2}e^{-\kappa_{\text{abs}}(f)D_d},
\end{equation}
where $D_d = \|\mathbf{p}_r - \mathbf{p}_t\|$. Assuming far-field, the \ac{LoS} channel is rank-one. Then, analog beamforming and combining yield the receive \ac{SNR}
\begin{equation}\tag{24}\label{eq:snr_mMIMO}
\text{SNR}_{\text{MIMO}} = \frac{N_rN_tP_t\text{PL}_{\text{MIMO}}}{\sigma^2}.
\end{equation}
Lastly, the respective power consumption is calculated as\footnote{The power consumption of signal processing is neglected.} 
\begin{equation}\tag{25}
P^{\text{MIMO}}_c = P_t + N_r(P_{\text{PS}} + P_{\text{PA}}) + N_t(P_{\text{PS}} + P_{\text{PA}} ),  
\end{equation}
where $P_{\text{PS}} = 42$ mW and $P_{\text{PA}} = 60$ mW are the power consumption values for a phase shifter and a power amplifier at $f = 300$~GHz, respectively~\cite{aosa_thz}.
\begin{figure}[t]
	\centering
	\includegraphics[width=1.08\linewidth]{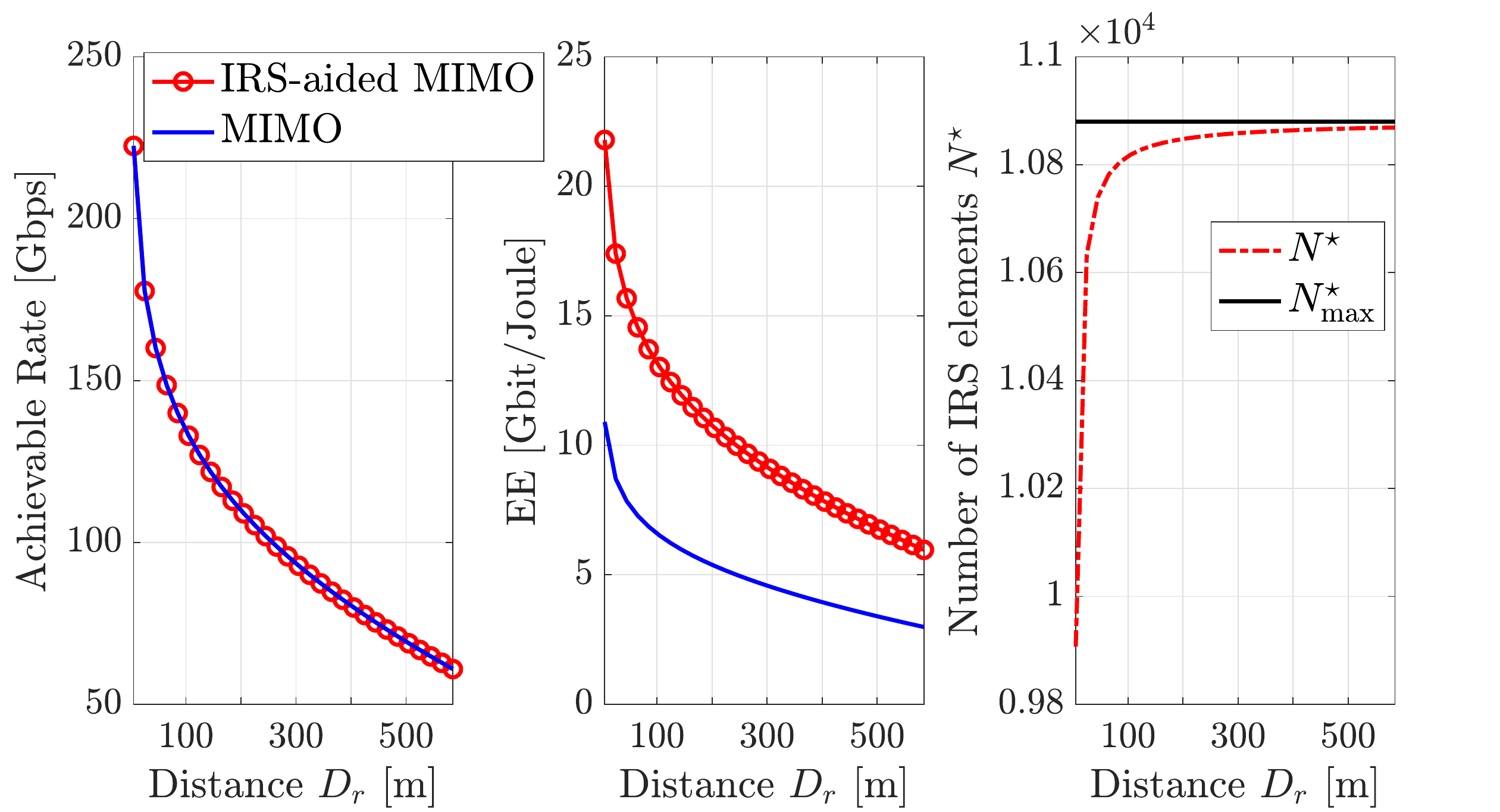}
	\caption{Achievable rate, \ac{EE}, and number of IRS elements versus distance $D_r$ for $\alpha=2$ and a fixed IRS location at (0,0,0). In the MIMO system, $N_t = 100$ and $N_r = 100$. The other parameters are $G_t = G_r = 20$~dBi, $P_t = 10$ dBm, $\sigma^2 = -174$ dBm/Hz, $B=10$~GHz, $f = 300$ GHz, $L_x=L_y=\lambda/2$, $\mathbf{p}_t = (0,-0.6,1)$ with $D_t = 1.16$ m, and $\mathbf{p}_r = (0, D_r,1)$.}
	\label{fig:Fig_EE}
\end{figure}
\subsection{IRS-Assisted MIMO System}
The Tx and Rx perform beamforming and combining to communicate a single stream through an \ac{IRS} of $N$ elements. Due to the directional transmissions, the Tx-Rx link is negligible, and thus is ignored. The received signal through the Tx-IRS-Rx channel is given by
\begin{equation}\tag{26}
y  = \mathbf{w}^H(\mathbf{H}_r\mathbf{\Phi}\mathbf{H}_t \mathbf{f}s + \tilde{\mathbf{n}}), 
\end{equation}
where $\mathbf{w}\in\mathbb{C}^{N_r\times 1}$ is the combiner, $\mathbf{f}\in\mathbb{C}^{N_t\times 1}$ is the beamformer, $\mathbf{H}_t\in\mathbb{C}^{N\times N_t}$ is the channel from the Tx to the \ac{IRS}, $\mathbf{H}_r\in\mathbb{C}^{N_r\times N}$ is the channel from the \ac{IRS} to the Rx, and $\tilde{\mathbf{n}}\sim\mathcal{CN}(\mathbf{0},\sigma^2\mathbf{I}_{N_r})$ is the noise vector. For ease of exposition, we assume far-field for both the Tx and the Rx. Then, 
\begin{align}
\mathbf{H}_r &= \sqrt{\text{PL}_r}e^{-jkD_r} \mathbf{a}_r(\phi_{\text{rx}},\theta_{\text{rx}})\mathbf{a}^H_{\text{IRS}}(\phi_r,\theta_r),\tag{27}\\
\mathbf{H}_t &= \sqrt{\text{PL}_t}e^{-jkD_t} \mathbf{a}_{\text{IRS}}(\phi_t,\theta_t)\mathbf{a}^H_t(\phi_{\text{tx}},\theta_{\text{tx}}),\tag{28}
\end{align}
where $\text{PL}_t \approx \text{PL}^t_{n,m}$ and $\text{PL}_r \approx \text{PL}^r_{n,m}$; the far-field response vectors $\mathbf{a}_r(\cdot,\cdot)$, $\mathbf{a}_t(\cdot,\cdot)$, and $\mathbf{a}_{\text{IRS}}(\cdot,\cdot)$ are specified according to Remark~\ref{remark1}. For $\mathbf{f} = \mathbf{a}_t(\phi_{\text{tx}},\theta_{\text{tx}})/\sqrt{N_t}$, $\mathbf{w}^H=\mathbf{a}^H_r(\phi_{\text{rx}},\theta_{\text{rx}})/\sqrt{N_r}$, and proper $\mathbf{\Phi}$, the receive \ac{SNR} is
\begin{equation}\tag{29}\label{eq:snr_IRS}
\text{SNR}_{\text{IRS}} = \frac{N_tN_rN^2P_t\text{PL}_{\text{IRS}}}{\sigma^2},
\end{equation}
where $\text{PL}_{\text{IRS}}$ is the path loss~\eqref{eq:PL_nm} of the \ac{IRS}-aided link. Using varactor diodes, the power expenditure of an \ac{IRS} element is negligible~\cite{nf_pathloss_model2}. Thus, the power consumption is determined as
\begin{equation}\tag{30}
P^{\text{IRS}}_c(N_t,N_r)  = P_t + N_r(P_{\text{PS}} + P_{\text{PA}}) + N_t(P_{\text{PS}} + P_{\text{PA}} ).
\end{equation}
\begin{proposition}\label{prop1}
The \ac{IRS}-aided system with $N_t/\alpha$ and $N_r/\alpha$ attains a higher \ac{SNR} than MIMO with $N_t$ and $N_r$ for 
\begin{equation}\tag{31}\label{eq:opt_number}
N^{\star} \geq \alpha\frac{\lambda}{L_xL_y}\frac{D_tD_r}{\sqrt{F(\theta_t,\phi_r,\theta_r)}D_d}e^{-\frac{1}{2}\kappa_{\emph{abs}}(f)(D_d - D_r-D_t)}.
\end{equation} 
\begin{proof}
According to~\eqref{eq:snr_mMIMO} and~\eqref{eq:snr_IRS}, the \ac{IRS}-aided system attains a higher SNR for $N^{\star} \geq \sqrt{\alpha^2\text{PL}_{\text{MIMO}}/\text{PL}_{\text{IRS}}}$, which gives the desired result after basic algebra.
\end{proof}
\end{proposition}
Using Proposition~\ref{prop1}, we can now decrease the number of Tx and Rx antennas by a factor $\alpha$ to reduce the power consumption as
\begin{align}
P^{\text{IRS}}_c(N_t/\alpha,N_r/\alpha) &= P_t + \frac{N_r}{\alpha}(P_{\text{PS}} + P_{\text{PA}}) + \frac{N_t}{\alpha}(P_{\text{PS}} + P_{\text{PA}}) \nonumber\\ 
&\approx P^{\text{MIMO}}_c/\alpha,\tag{32}
\end{align}
while keeping the achievable rate fixed. Hence, the \ac{EE} gain with respect to \ac{MIMO} is approximately equal to $\alpha$. 
\begin{figure}[t]
	\centering
	\includegraphics[width=0.72\linewidth]{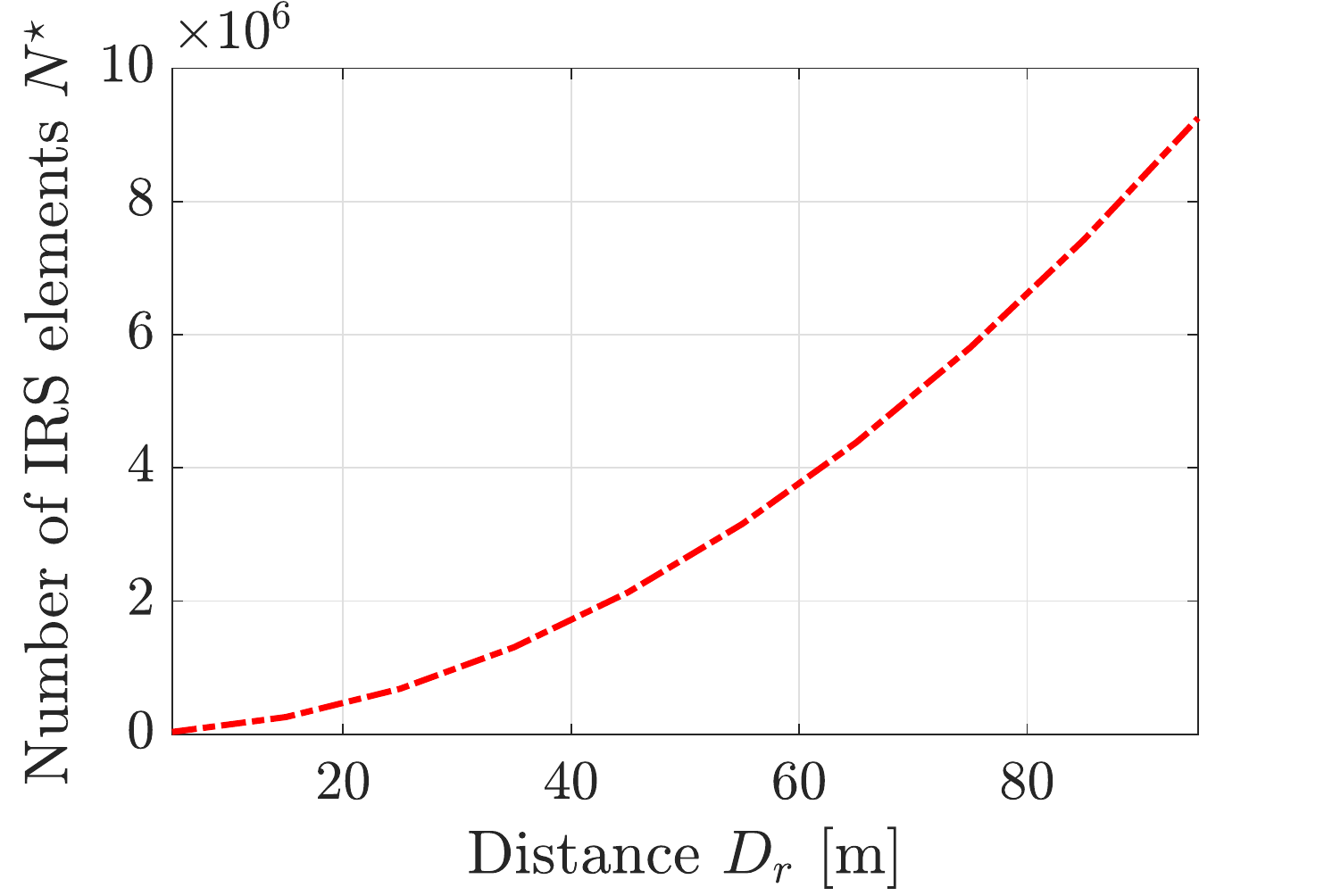}
	\caption{Number of IRS elements $N^{\star}$ versus distance $D_r$ for $\alpha=2$; $\mathbf{p}_t = (0,-0.6,1)$, $\mathbf{p}_r = (0, D_r,1)$, IRS at $(0,(D_r - y_t)/2,1)$ with $y_t = -0.6$, $f=300$ GHz, $\kappa_{\text{abs}}(f)=0.0033$ m$^{-1}$, and $L_x=L_y=\lambda/2$.}
	\label{fig:Fig_Nelements}
\end{figure}

\section{Numerical Results}
In this section, we assess the performance of IRS-aided \ac{THz} communication through numerical simulations. For this purpose, we calculate the achievable rate as
\begin{equation}\tag{33}
R = B\log_2(1 + \text{SNR}),
\end{equation}
where $B$ is the signal bandwidth. Moreover, the \ac{EE} is specified as $\text{EE} \triangleq R/P_c$. 
\subsection{Energy Efficiency}
We consider a MIMO setup with $N_t=N_r=100$ antennas, i.e., $10\times 10$-element planar arrays. From Fig.~\ref{fig:Fig_EE}, we verify that the IRS-assisted system with $N_t=N_r = 50$ antennas offers a two-fold \ac{EE} gain. Consequently, an \ac{IRS} can provide an alternative communication link, in addition to~\ac{LoS}, where the Tx and Rx employ a smaller number of antennas to communicate with each other, hence saving energy. Note, though, that the suggested benefits are valid when: 1) the power expenditure of \ac{IRS} elements is negligible compared to that of conventional phase shifters; 2) the Tx operates near the \ac{IRS} in order to have a reasonable number of reflecting elements $N^{\star}$; and 3) reflection losses are small~\cite{aperture_eff}. 

\subsection{IRS Placement and  Near-Field Beamfocusing}
We now investigate the impact of the \ac{IRS} position on the number of \ac{IRS} elements $N^{\star}$. For the deployment in Fig.~\ref{fig:Fig_EE}, $D_t$ is small, and hence $D^2_r \approx D^2_t + D^2_d$. Further, $\phi_r = \pi/2$ which gives $F(\theta_t,\phi_r,\theta_r) = \cos^2\theta_t$. Then,~\eqref{eq:opt_number} reduces to 
\begin{equation}\tag{34}
N^{\star} = \alpha\frac{\lambda}{L_xL_y}\frac{D_tD_r}{\cos\theta_t\sqrt{D^2_r  - D^2_t }}e^{-\frac{1}{2}\kappa_{\text{abs}}(f)(\sqrt{D^2_r  - D^2_t} - D_r-D_t)},
\end{equation}
which takes the asymptotic value 
\begin{equation}\tag{35}
N_{\max}^{\star} = \alpha\frac{\lambda}{L_xL_y}\frac{D_t}{\cos\theta_t}e^{\frac{1}{2}\kappa_{\text{abs}}(f)D_t}
\end{equation}
as $D_r\to \infty$; this follows from $\sqrt{D^2_r  - D^2_t}\approx D_r$ for $D_r\gg~D_t$. Thus, $N^{\star}$ is bounded for a fixed IRS position near the Tx. Due to symmetry, the same holds when the IRS is near the Rx. For instance, $N^{\star}_{\max} = 10,880$ in Fig.~\ref{fig:Fig_EE}.  In contrast, when the \ac{IRS} is deployed always in the middle of the Tx and Rx, $N^{\star}$ increases as $O(D_tD_r)$. This scaling law is depicted in Fig.~\ref{fig:Fig_Nelements}. Consequently, the \ac{IRS} has to be close to the link ends in order to compensate for the severe propagation losses with a practical number of reflecting elements. Note that similar findings were reported in~\cite{tutorial}. In this case, the Tx/Rx will be in the Fresnel zone of the \ac{IRS} where \text{near-field beamfocusing} becomes the optimal processing strategy; otherwise, the \ac{EE} gains previously discussed cannot be attained.

\begin{figure*}[t]
	\begin{align}\label{eq:losses}
	G &= \frac{\left|\sum_{n=0}^{N_x-1} e^{-jk \frac{(nd_x)^2(1-\cos^2\phi_t\sin^2\theta_t)}{2D_t}} \right|^2}{N_x^2}\frac{\left|\sum_{m=0}^{N_y-1}e^{-jk\frac{(md_y)^2(1-\sin^2\phi_t\sin^2\theta_t)}{2D_t}}\right|^2}{N^2_y} \nonumber\\
	&\approx \frac{\left|\sum_{n=0}^{N_x^2-1} e^{-jk \frac{nd_x^2(1-\cos^2\phi_t\sin^2\theta_t)}{2D_t}} \right|^2}{N_x^4}\frac{\left|\sum_{m=0}^{N_y^2-1} e^{-jk\frac{md_y^2(1-\sin^2\phi_t\sin^2\theta_t)}{2D_t}}\right|^2}{N_y^4} \nonumber\\
	&= \left|D_{N_x^2}\left(\frac{2\pi}{\lambda} \frac{d_x^2(1-\cos^2\phi_t\sin^2\theta_t)}{2D_t}\right)\right|^2  \left|D_{N_y^2}\left(\frac{2\pi}{\lambda}\frac{d_y^2(1-\sin^2\phi_t\sin^2\theta_t)}{2D_t}\right)\right|^2.\tag{22}
	\end{align}
	\hrulefill
\end{figure*}
\section{Conclusions and Future Work}
We studied the channel modeling and performance of \ac{IRS}-assisted \ac{THz} communication. First, we introduced a spherical wave channel model and employed plate scattering theory to derive the path loss. We next showed that the path loss is nearly constant across the \ac{IRS} thanks to its small physical size. However, due to the large number of reflecting elements with respect to the wavelength, the Fresnel zone of a THz \ac{IRS} is of several meters. To this end, we analyzed the power gain under near-field beamfocusing and conventional beamforming, and proved the suboptimality of the latter. One implication of this is that the \ac{IRS} needs to know the exact location of the Tx and/or Rx, rather than their angular information, to perform beamfocusing. Capitalizing on the derived model, we finally investigated the \ac{EE} scaling law of \ac{IRS}-aided \ac{MIMO}, and showed that it can outperform \ac{MIMO}. Numerical results consolidate the potential of \ac{IRS}s for \ac{THz} communication. For future work, it would be interesting to study the reflection matrix design for a multi-antenna Tx/Rx that operates in the Fresnel zone of the \ac{IRS}, as well as pursue an \ac{EE} analysis under hardware impairments and channel estimation overheads.

\end{document}